\documentclass[sigconf]{acmart} 
\AtBeginDocument{%
  }

\copyrightyear{2025}
\acmYear{2025}
\setcopyright{rightsretained}
\acmConference[CHI '25]{CHI Conference on Human Factors in Computing Systems}{April 26-May 1, 2025}{Yokohama, Japan}
\acmBooktitle{CHI Conference on Human Factors in Computing Systems (CHI '25), April 26-May 1, 2025, Yokohama, Japan}\acmDOI{10.1145/3706598.3714154}
\acmISBN{979-8-4007-1394-1/25/04}

\makeatletter
\gdef\@copyrightpermission{
  \begin{minipage}{0.3\columnwidth}
   \href{https://creativecommons.org/licenses/by-nc-nd/4.0/}{\includegraphics[width=0.90\textwidth]{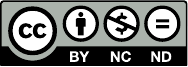}}
  \end{minipage}\hfill
  \begin{minipage}{0.7\columnwidth}
   \href{https://creativecommons.org/licenses/by-nc-nd/4.0/}{This work is licensed under a Creative Commons Attribution-NonCommercial-NoDerivs International 4.0 License.}
  \end{minipage}
  \vspace{5pt}
}
\makeatother

\usepackage{xspace}

\usepackage{acmart-taps}
\usepackage{tikz}
\usepackage{xcolor,colortbl}
\newcommand\mybox[2][]{\tikz[overlay]\node[fill=blue!20,inner sep=2pt, anchor=text, rectangle, rounded corners=1mm,#1] {#2\xspace};\phantom{ #2}}

\newcommand{\loz}[2][blue]{~\mybox[fill=#1!20]{\textsc{#2}}}

\aptLtoXcmd{\newcommand{\tryButton}[0]{\colorbox[HTML]{ccccff}{Try}\xspace}}
{\newcommand{\tryButton}[0]{ \loz[blue]{Try}\xspace}}

\aptLtoXcmd{\newcommand{\revertButton}[0]{\colorbox[HTML]{ffe5cc}{Revert}\xspace}}
{\newcommand{\revertButton}[0]{ \loz[orange]{Revert}\xspace}}

\aptLtoXcmd{\newcommand{\dqSec}[0]{\colorbox[HTML]{ccffcc}{DQs}\xspace}}
{\newcommand{\dqSec}[0]{ \loz[green]{DQs}\xspace}}

\aptLtoXcmd{\newcommand{\crSec}[0]{\colorbox[HTML]{ffcccc}{Reqs}\xspace}}
{\newcommand{\crSec}[0]{ \loz[red]{Reqs}\xspace}}

\aptLtoXcmd{\newcommand{\idSec}[0]{\colorbox[HTML]{f2ccd9}{Decs}\xspace}}
{\newcommand{\idSec}[0]{ \loz[purple]{Decs}\xspace}}

\aptLtoXcmd{\newcommand{\uaSec}[0]{\colorbox[HTML]{cce5e5}{UAbs}\xspace}}
{\newcommand{\uaSec}[0]{ \loz[teal]{UAbs}\xspace}}

\aptLtoXcmd{\newcommand{\dg}[1]{\colorbox[HTML]{cce5cc}{DG{#1}}}}
{\newcommand{\dg}[1]{\loz[darkgreen]{DG{#1}}}}

\definecolor{darkgreen}{rgb}{0.8,0.8,0.8}
\usepackage{multirow}
\usepackage{makecell}

\newcommand{\keyIdea}[1]{#1}

\definecolor{darkgreen}{rgb}{0,0.5,0}

\newcommand{\edits}[1]{#1}
\newcommand{\deletes}[1]{}

\usepackage{graphicx,subcaption} 
\begin{document}

\title[Beyond Code Generation: LLM-supported Exploration of the Program Design Space]{Beyond Code Generation:\\LLM-supported Exploration of the Program Design Space}

\author{J.D. Zamfirescu-Pereira}
\email{zamfi@berkeley.edu}
\affiliation{%
  \institution{UC Berkeley}
  \city{Berkeley}
  \state{CA}
  \country{USA}
}

\author{Eunice Jun}
\email{emjun@cs.ucla.edu}
\affiliation{%
  \institution{UCLA}
  \city{Los Angeles}
  \state{CA}
  \country{USA}
}

\author{Michael Terry}
\email{michaelterry@google.com}
\affiliation{%
  \institution{Google DeepMind}
  \city{Cambridge}
  \state{MA}
  \country{USA}
}

\author{Qian Yang}
\email{qianyang@cornell.edu}
\affiliation{%
  \institution{Cornell University}
  \city{Ithaca}
  \state{NY}
  \country{USA}
}

\author{Bj\"orn Hartmann}
\email{bjoern@eecs.berkeley.edu}
\affiliation{%
  \institution{UC Berkeley}
  \city{Berkeley}
  \state{CA}
  \country{USA}
}

\begin{abstract}
In this work, we explore explicit Large Language Model (LLM)-powered support for the iterative design of computer programs. Program design, like other design activity, is characterized by navigating a space of alternative problem formulations and associated solutions in an iterative fashion. LLMs are potentially powerful tools in helping this exploration; however, by default, code-generation LLMs deliver code that represents a particular point solution. This obscures the larger space of possible alternatives, many of which might be preferable to the LLM’s default interpretation and its generated code. We contribute an IDE that supports program design through generating and showing new ways to frame problems alongside alternative solutions, tracking design decisions, and identifying implicit decisions made by either the programmer or the LLM. 
In a user study, we find that with our IDE, users combine and parallelize design phases to explore a broader design space---but also struggle to keep up with LLM-originated changes to code and other information overload. These findings suggest a core challenge for future IDEs that support program design through higher-level instructions given to LLM-based agents: carefully managing attention and deciding what information agents should surface to program designers and when.

\end{abstract}


\begin{teaserfigure}
  \centering
  \includegraphics[width=\textwidth]{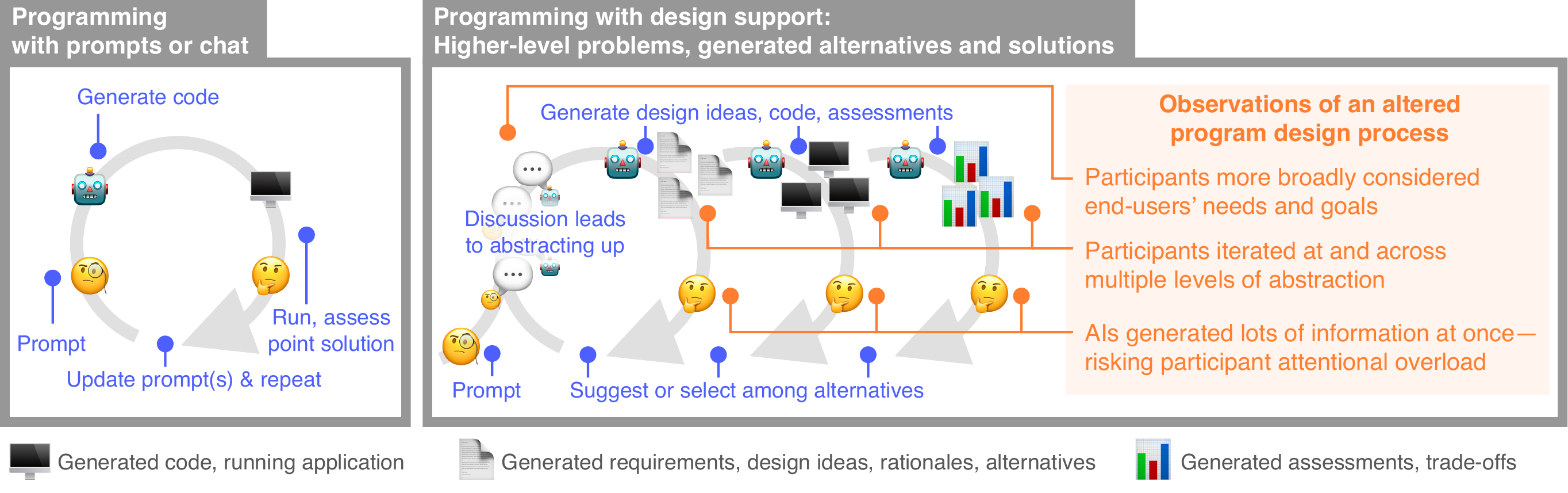}
  \caption{Today, LLM-powered programming assistance resembles a repeating cycle of  ``prompt''-to-``code'' (left) by re-running generation with every prompt change. {\textsc{Pail}\xspace}, our IDE (right), instead encourages developers to abstract up, pulling them towards a deeper understanding of the problem space through an exploration of alternative problem formulations and solutions. Throughout this exploration, {\textsc{Pail}\xspace} tracks design goals, changing requirements, and surfaces implicit decisions---but the breadth and depth of information {\textsc{Pail}\xspace} generates can be overwhelming.}
  \Description{}
  \label{fig:teaser}
  \end{teaserfigure}

\maketitle

\section{Introduction}

A common vision of future computer programming relies on Large Language Models (LLMs) to do it all: identify requirements, write code and tests, perform simulated QA testing, deploy, etc. We are making rapid progress towards this vision on many fronts~\cite{hong_metagpt_2023,di_fede_idea_2022,noauthor_artifacts_2024,chen_evaluating_2021,si_design2code_2024}. Yet the role of the human in this process is under-explored. In the predominant view, humans will be limited to setting an initial goal, possibly clarifying a few questions asked, and then rating the output. For some changes to the goal or generated code, the process will restart from, approximately, scratch. Many versions of this vision exist: some scale up the ``autocomplete,'' CoPilot-like~\cite{github_github_2021} interactions relying on recognition-over-recall~\cite{nielsen_10_nodate}, others scale up the ``chat''~\cite{noauthor_artifacts_2024} or focus on providing point solutions~\cite{hong_metagpt_2023}.

While these visions best suit an understanding of program design as starting with one known goal, program design in fact involves a design process iterating on both the implementation details and design goals. For instance, a journalist may begin an analysis with the goal of assessing a relationship between two variables but then in the process of exploring possible visualizations realize that one of the variables has a lot of missing data. As a result, the journalist can pivot to assess another pair of relationships or consider alternative visualizations and statistical analyses to triangulate the initial relationship of interest. Similarly, a game creator faces questions of narrative, characters, interactions. Even a straightforward backend engineering task (e.g., storage system selection) requires consideration of trade-offs (e.g., latency, consistency, scale) that ultimately require refining the higher-level goals of the task (e.g., terms of a latency service agreement).

\begin{figure*}
  \centering
  \includegraphics[width=.7\linewidth]{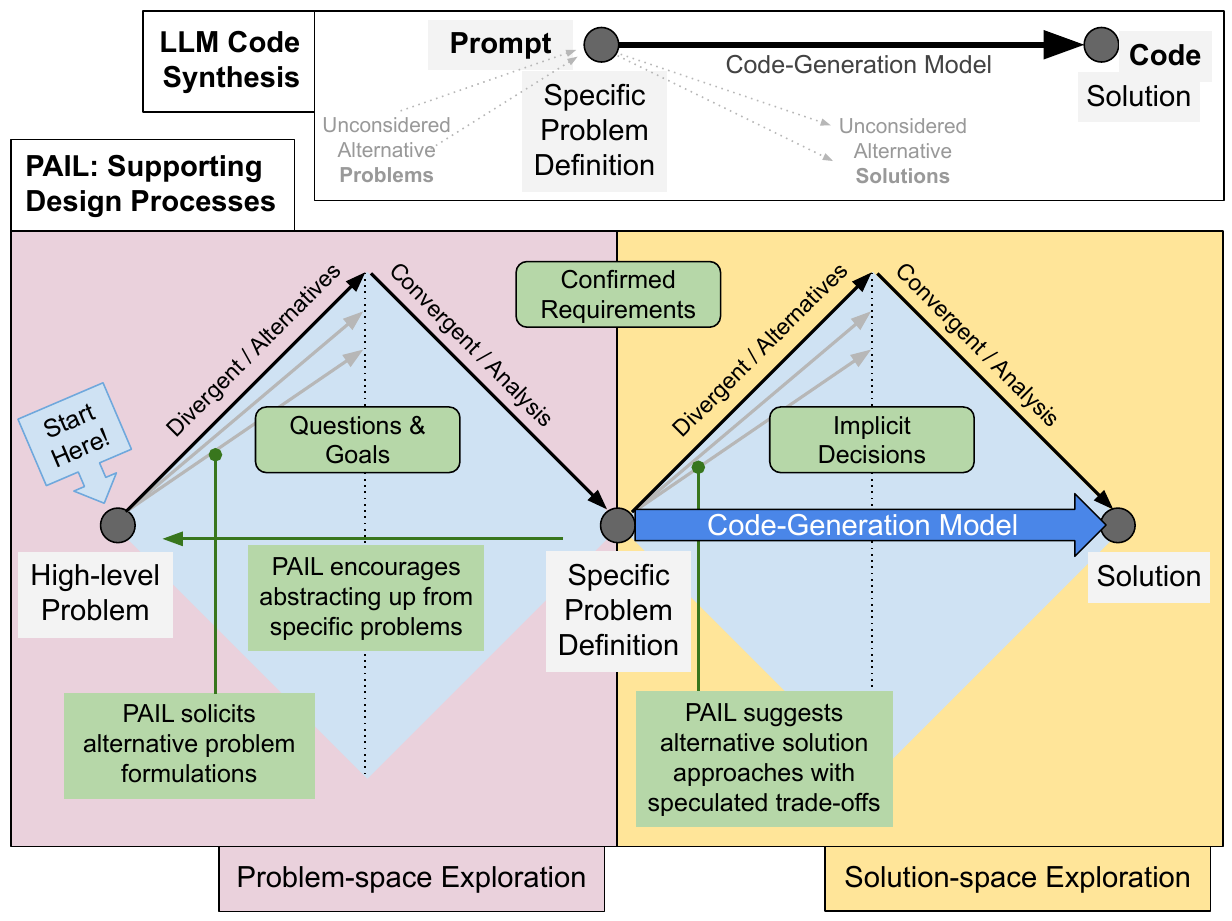}
  \caption{{\textsc{Pail}\xspace}'s capabilities overlaid on the UK Design Council's Double Diamond~\cite{design_council_framework_2004}. Building beyond a ``prompt''-to-``code'' view of LLM code generation (top), {\textsc{Pail}\xspace} (bottom) embeds this capability within an IDE that supports developers through a larger design process: by abstracting up from a specific problem to a higher-level one, exploring alternative problems and alternative solutions, tracking design goals and requirements, and surfacing implicit decisions (green boxes).}
  \Description{}
  \label{fig:diamonds}
\end{figure*}

Viewing programming as a design activity illuminates the human's need for tighter iterative loops and more fine-grained control than implied by the predominant vision. Design activity begins with consideration of different \textit{problem} formulations before solutions are even discussed---as captured iconically by the \textit{Double Diamond} (see, e.g.,~\cite{design_council_framework_2004}), which we adapt to illustrate our targeted capabilities, in Figure~\ref{fig:diamonds}. Designers test high-level hypotheses (e.g., around user needs, or problem urgency) with sketches and prototypes before expending engineering effort. All these explorations feed back into a better understanding of the problem and solution design spaces---but take time, and are often limited by human availability and cost. Temptingly, LLMs promise a near-free pool of ``cognitive resources'' that could in theory complement human cognition to improve the experience of designing and programming interactive artifacts and improve the final artifacts' overall quality---by using LLMs to complement human limitations on time and cost, and using humans to overcome LLMs' limitations on domain expertise and end-user understanding. We explore this promise in this work.

Through a formative study of creating small interactive sketches using ChatGPT\footnote{\url{https://www.chatgpt.com/}} and Claude,\footnote{\url{https://www.claude.ai/}} we find that (1) working through a user-centered design process in chat alone leads to lost requirements and a lot of scrolling through to find prior questions, decisions, and discussions, and (2) individual point solutions---without consideration of alternatives---lead to significant anchoring bias. Combining these findings with knowledge from prior work in AI assistance for design, we develop the {\textsc{Pail}\xspace} IDE. 

{\textsc{Pail}\xspace} embeds a chat agent with two other LLM-based agents into an integrated environment to facilitate higher-level consideration of design choices while programming with LLMs, enabling users to \textbf{design} and \textbf{iterate} on programs while working at higher levels of abstraction than the code itself. At the highest level, {\textsc{Pail}\xspace} elicits design alternatives, rationales, and prompts to guide users toward user-centered design principles. {\textsc{Pail}\xspace} also speculatively proposes and explores alternative problem and solution formulations, generating interactive prototypes to support epistemic goals through testing and other assessment. At the most granular level, {\textsc{Pail}\xspace} tracks requirements and implicit decisions made by the (black-box) LLM without user input, surfacing abstractions to establish common ground and clarify understanding. As an IDE, {\textsc{Pail}\xspace} also allows users to directly edit the code itself, providing full control over implementation details.

Through a lab study with 11 participants, we use {\textsc{Pail}\xspace} as a probe to understand the benefits, limitations, and some future challenges for LLM-powered programming tools' support for program design. {\textsc{Pail}\xspace} helps participants consider their audience and communication goals; participants express appreciation for having a direct summary and for the ability to manipulate those decisions in situ, and 100\% of participants find at least one unconsidered alternative that influenced their design work. Further, we find that rapid code generation allows actual executable interactive programs to fit into a \textit{sketch} role: a cheap, ``disposable,''~\cite{buxton_sketching_2010} epistemological artifact, rather than a \textit{work-in-progress} prototype representing an investment of time and resources---an ability that is suggestive of an altered program design process. Finally, our findings suggest an emerging issue in managing user attention: as more agents and UI affordances lay claim to being helpful and are integrated into developer environments and workflows, these tools will also need to competently balance user attentional capacities and desire for agency. We discuss these observations' implications for future tools supporting programming and other design activities, and suggest opportunities for further research.

This paper makes three contributions: First, it describes {\textsc{Pail}\xspace}, a prototype IDE that introduces a new set of interactions extending chat to support program design activity at higher levels of abstraction. Second, it offers a rich description of how and when {\textsc{Pail}\xspace}'s design support is useful (and not) for creating interactive software prototypes. Lastly, it identifies core upcoming challenges for a future of programming in which programmers relay abstract instructions to LLM agents that synthesize code: these agents will, critically, need to manage and direct user attention, keeping users abreast of AI-initiated changes and carefully considering what information to show.

\section{Related Work}
Our work draws on empirical studies and theories of design and recent work on LLMs for creative tasks. Here, we review prior work on systems for sketching and comparing alternatives within design processes; the nascent area of LLM-powered programming assistance as well as pre-LLM approaches; and, lastly, approaches to making better use of LLMs in structured complex tasks.

\subsection{Program Design Processes; Sketching and Comparing Alternatives} \label{sec:rw-alternatives}

In the fields of design and HCI, the process of designing is characterized by exploring the space of solutions and iteratively reformulating the problem to solve. Sketching, a form of rapid prototyping for quickly exploring the essential dimensions of a solution space, is a widely accepted, studied, and practiced tool for thought~\cite{buxton_sketching_2010, landay_interactive_1995}---serving in part to help designers explore alternatives, sometimes even in parallel~\cite{dow_parallel_2010}. Described by Buxton as distinct from \textit{prototyping}, \textit{sketching} aims to evoke, rather than describe. A prototype is often constructed for a specific goal, such as to explore how an artifact might work, look and feel, or relate to other components of a system~\cite{houde_what_1997}, and can be considered a ``work in progress'' in the sense that it is meant to faithfully reproduce that aspect of a final artifact, perhaps even to confirm a particular design decision, and as such represents an investment of effort (and potentially materials). A sketch, in contrast, is meant to be quick, disposable and plentiful, suggestive, and serves more of a ``cognitive offloading'' role to build a shared understanding~\cite{buxton_sketching_2010}.

In divergent, exploratory phases of the program design process, programmers also often engage in \textit{Exploratory Programming}~\cite{beth_kery_exploring_2017,sheil_datamation_1986}, writing code as a \textit{medium to prototype} with different ideas, with an open-ended goal and no existing specification. In such a phase, programmers may engage in \textit{Opportunistic Programming}~\cite{brandt_two_2009,brandt_writing_2009}, a form of programming in which speed and ease are prioritized over robustness, maintainability, or other engineering goals. Copied code found on the web is a common hallmark---and today's LLMs' code generation capabilities provide a new venue for finding code to incorporate into a prototype.

Considering and examining alternatives, in these phases an explicit activity forming part of the design process, has a long history in HCI too: recent examinations in the context of prompting text-to-image models like \textsc{Promptify}~\cite{brade_promptify_2023} and \textsc{Dreamsheets}~\cite{almeda_prompting_2024} echo prior systems like Kery et al's \textsc{Variolite} and \textsc{Verdant} for computational notebooks~\cite{kery_variolite_2017, kery_towards_2019}, Hartmann et al's \textsc{Juxtapose} system for exploring alternatives in parallel in code~\cite{hartmann_design_2008}, and Terry et al.'s \textit{Parallel Paths} approach for parameterized vector art~\cite{terry_variation_2004}. These systems in turn trace their lineages back to Marks et al's Design Galleries~\cite{marks_design_1997} for expensive-to-generate computer images, and are complemented by a set of theory-driven work aimed at understanding how and why these processes help designers~\cite{lunzer_subjunctive_2008, rawn_understanding_2023}. 
These systems and studies all point to parallel exploration and tracking of alternatives as a well-established, essential process in exploratory programming~\cite{kery_designing_2017}.

\subsection{AI Assistance for Programming and Design}

Our work here also builds on two major threads of research in AI assistance: assistance for programming, and assistance for design. On the programming side, the biggest impact comes from autocomplete-focused assistance through tools like CoPilot~\cite{github_github_2021}, which has been extensively studied~\cite{barke_grounded_2023,prather_its_2023,denny_conversing_2023} and shown to serve both ``recognition-over-recall''~\cite{nielsen_10_nodate} and epistemic (e.g., ``oh, I didn't know about list comprehension in Python!'') goals~\cite{barke_grounded_2023}. Other work has explored how LLMs can be used in the service of design processes for creative coding~\cite{angert_spellburst_2023}, data analysis~\cite{mcnutt_design_2023,liu_what_2023,lam_concept_2024,gu_how_2024}, and even, circularly, for the assessment of LLM outputs themselves~\cite{shankar_who_2024}.

Pre-LLM work in both AI assistance~\cite{chen_towards_2016} and the usability of code synthesis~\cite{jayagopal_exploring_2022,lubin_program_2019} is also relevant here, offering insight into the kinds of assistance programmers are looking for: support for writing mundane boilerplate or glue code, for reasoning, and for rapid iteration. Traditional code synthesis has found uses in a number of ways, most notably for HCI through a line of work on programming by demonstration---often repeated demonstrations following a feedback-driven design process, in the service of spreadsheet formula construction~\cite{gulwani_automating_2011} and web scraping and automation~\cite{li_sugilite_2017,barman_ringer_2016}.

This empirical work is complemented by a set of theoretical, speculative, and design-oriented work around the opportunities and challenges of effectively instructing AI systems and designing both with and through them. These include studies of prompting~\cite{zamfirescu-pereira_herding_2023,zamfirescu-pereira_why_2023,liu_design_2022}, impacts of current LLMs on creative design processes~\cite{anderson_homogenization_2024,suh_sensecape_2023} speculative future design processes~\cite{vaithilingam_imagining_2024}, new conceptual models (e.g.,~\cite{terry_ai_2023,subramonyam_bridging_2024}), and questions of agency (e.g.,~\cite{lawton_when_2023}) and perception (e.g.,~\cite{khadpe_conceptual_2020}). This body of work emphasizes the extent to which iteration is critical to design, especially when working with black-box models, that evaluating LLM outputs for correctness is challenging due to intrinsic unpredictability, and that steering LLMs benefits from understanding how LLMs go about performing the tasks a user asks them to perform.

\subsection{Workflows Integrating LLMs in Complex Tasks}

In order to take advantage of LLMs' capabilities to perform simple tasks well and apply them to more complex tasks, researchers have explored workflows that integrate LLMs. Wu et al. proposed chaining as a technique for connecting LLM calls to each other~\cite{wu_ai_2022}. Building on this interaction model, Arawjo et al. developed ChainForge~\cite{arawjo_chainforge_2023}, an interface for composing LLM prompts and assessing the results of prompts. Exploring alternatives to linear composition, Kazemitabaar et al.~\cite{kazemitabaar_improving_2024} compare two different forms of task composition: (i) phase-wise decomposition which batches steps together and (ii) step-wise decomposition which iterates on each step piecewise. Across this work, a common finding is that systems need to scaffold LLM usage in order for users \textit{across experience levels} to make the most of an LLM's capabilities.

This set of work also shows us that individual interventions to aid in design, in organization, in evaluation, and in grounding assumptions can all be helpful. Here, we aim to shed some light on what challenges will arise when we begin to build more complex cognitive support tools that integrate multiple of these affordances into a single system. Indeed, research in the AI community has shifted towards developing cognitive architectures for coordinating multiple agents~\cite{sumers_cognitive_2024}, even to complement each other's strengths and weaknesses~\cite{hong_metagpt_2023}.


\section{Designing {\textsc{Pail}\xspace}}

Through this work, we aim to understand how explicit design support can impact programmers' design and prototyping processes. We adopt a Research through Design~\cite{zimmerman_research_2007} approach and develop {\textsc{Pail}\xspace}, a web-based IDE for programming \texttt{p5.js} sketches, to probe into what programming with LLMs could look like in the future. 

\subsection{Design Process}
We developed {\textsc{Pail}\xspace} iteratively based on the described prior work and a formative study. Throughout our design process, we discussed and incorporated feedback among the co-authors.

Two practical considerations constrained our tool and study design spaces upfront; these led us to focus primarily on experienced programmers making small interactive prototypes for personal or (non-deployment) professional use:

First, the current fidelity of LLM code synthesis limits the complexity of the programs that can be effectively constructed or iterated upon by these models with limited human intervention---but we expect new models to continue to develop these capabilities. To avoid having our observations unduly influenced by today's model failure modes, we target smaller programs that can be fully included in LLM context windows, in a language (JavaScript) and using a framework (\texttt{p5.js}) that are well-represented in LLMs' training data.

Second, we are interested specifically in \textit{design support}, not necessarily in supporting end-users learning how to program---there is plenty of great work in that area already. Thus, we primarily target participants with a strong understanding of programming.

\subsection{Formative Study}

We conducted a formative study to identify challenges that emerge when following common design processes for program design using an LLM-powered chatbot. We recruited five participants from within the HCI and Design-focused groups at our institution, with a range of experiences creating interactive software prototypes.  All participants were students in their final two years of undergraduate education or doctoral students.

Over each of six weeks, a subset of participants used ChatGPT or Claude to produce \texttt{p5.js} code prototypes for a variety of interactive programs, running those prototypes by copy-and-pasting any generated code into a standard \texttt{p5.js} environment.\footnote{\url{https://editor.p5js.org}} We chose ChatGPT and Claude for our study rather than GitHub CoPilot or other AI copilots because the latter tools do not provide design guidance nor engage in requirement elicitation, and require users to engage with code rather than higher-level abstractions.

Our pariticipants' interactive programs ranged from basic games and simulations (e.g., a fashion simulator that lets you try on---virtually---images of clothing) to showcases for artwork, aggregators for real estate listings, and other sensemaking tasks. Occasionally, participants were asked to create an interactive artifact \textit{for someone else}, or asked to observe someone not related to the study using ChatGPT or Claude for the first time.

Our formative study participants observed that tracking the outcomes of design discussions was a major challenge, as these discussions rapidly disappeared into chat history and became very challenging to find or recall---and, as decisions receded into the chat history further, the models themselves were also less likely to consider them in future iterations of the design. Additionally, both ChatGPT and Claude limited their responses to one of (1) a single point solution paired with a description of that solution, without discussion of trade-offs or alternative solutions, or (2) several possible solutions, paired with a description, but not comparing solutions with each other or discussing trade-offs among them. 

Participants expressed an appreciation for being able to quickly generate code to test out ideas, but found it challenging to later identify those experiments or revert to the code used for them. Participants also reported feeling unsure about the code the LLMs generated, looking at it relatively rarely---the common iterative loop consisted of running the generated code and asking the LLM directly to fix any observed errors, rather than reviewing the code directly for errors. Conspicuously lacking was support for an understanding of the generated code, beyond the high-level overviews produced by the LLMs---overviews which lacked mention or explanation of critical decisions or assumptions made by the LLM. Even when these were provided, they were typically buried within paragraphs of overview, requiring a close read to find, and participants almost never read these explanations closely because of a low signal-to-noise ratio.

Lastly, those participants who mediated a program generation exercise for someone else reported spending substantial time identifying what needs that user had, and then synthesizing solutions for those needs. These activities required more concrete epistemic goals in prototyping.

\subsection{Design Goals} \label{sec:designGoals}

From our formative study, we distilled a set of four design goals.

\dg1: {\textsc{Pail}\xspace} should provide explicit support for generating, keeping track of, and comparing alternative designs. 

\dg2: {\textsc{Pail}\xspace} should extract and track task requirements and decisions explicitly, outside of the primary conversation dialogue, to keep these salient and visible to both the human designer and the LLM that is generating code. 

\dg3: {\textsc{Pail}\xspace} should take reasonable steps to help users avoid reading the program code, keeping them at their desired level of abstraction in communicating in terms of functionality and goals---including by surfacing, in natural language, decision points that exist only implicitly in the code. 

\dg4: {\textsc{Pail}\xspace} should help users consider the needs and goals of the ultimate users of the programs being designed.

We considered several possible designs for showcasing alternatives (\dg1) and version histories in early prototypes of {\textsc{Pail}\xspace}, including a Git-like ``commit'' structure inspired by Litt~\cite{litt_patchwork_2024}. We ultimately selected a lightweight ``pull'' model influenced by Kery et al's observation that, in data science workflows, ``versioning'' in the traditional software sense can be too heavyweight to be useful~\cite{kery_towards_2019}. We also decided that, as suggested by Lunzer et al.~\cite{lunzer_subjunctive_2008}, {\textsc{Pail}\xspace} should offer \textit{prospective} (speculative) comparisons of possible alternatives, not only retrospective comparisons of selected alternatives. As a result, we implemented {\textsc{Pail}\xspace} to suggest alternatives whenever possible (\dg1).

To help ensure that user communications at a high level of abstraction were likely to refer to concepts the LLM also had a handle on, and based on recent findings in cognitive science on the discovery and usage of abstractions in conversation, we incorporated an explicit mechanism for grounding specific jargon with a textual description (\dg3). For example, consider a puzzle game with a specific win condition: explicitly calling out ``Win Condition,'' with a description of what that phrase means, offers users both a specific name to refer to this concept, and confirmation that both the user and {\textsc{Pail}\xspace} are using that phrase to refer to the same sort of underlying concept. 

A second affordance aimed at a maintaining a level of abstraction higher than code is a set of individual code change summaries, in natural language text, that accompany any code change {\textsc{Pail}\xspace} suggests. Incorporating this explicit mechanism for tracking code changes at a higher level allows users to avoid the ``context switch'' of shifting their thinking mode from design considerations, to code, and back again.

Lastly, explicitly surfacing ``implicit'' decisions (necessarily) made by the LLM when it synthesizes code in response to an ambiguous request provides a mechanism for considering these decisions and generating possible alternatives (\dg2,\dg3). These decisions can be critical to a design, and include decisions about how to represent data, like user progress through an application, or how to operationalize certain constraints, like how to validate a ``win condition'' in a game. In our formative study, we found that unsurfaced decisions were discussed only when these decisions had a visible impact the user noticed and chose to inquire about, but constrained the future design space regardless---so we chose to make them explicit here.

\subsection{System Design}
\begin{figure*}
  \includegraphics[width=.75\linewidth]{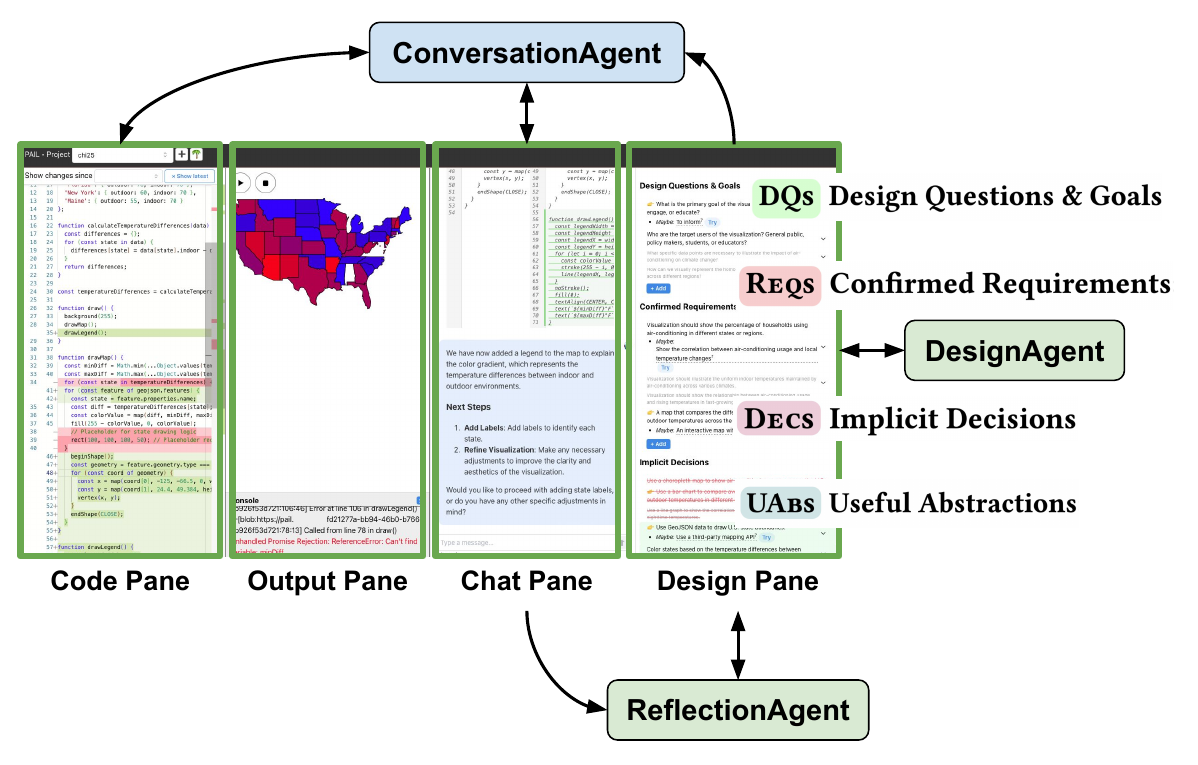}
  \caption{An overview of {\textsc{Pail}\xspace}'s user interface and related agents. {\textsc{Pail}\xspace} engages users in defining and refining (\dqSec) design goals, including target audience and desired impact, while tracking (\crSec) confirmed requirements and (\idSec) decisions implicit in an LLM's synthesized code, surfacing (\uaSec) useful abstractions, and offering explanations and alternatives across the program design space. A full-scale reproduction of the design pane's four views appears in Figure~\ref{fig:design-goals-snapshot}.}
  \Description{A snapshot of the design panel.}
    \label{fig:inset}
  \label{fig:agents}
\end{figure*}

What emerged from our design process was the need for two sets of entities: a set of ``agents'' that engage with the user directly (such as the {\textsc{ConversationAgent}\xspace}) or indirectly through UI affordances (such as the {\textsc{DesignAgent}\xspace} and {\textsc{ReflectionsAgent}\xspace}). These agents are complemented with a set of ``views'' (see below) aggregated into a ``design panel'' on the right-hand side of {\textsc{Pail}\xspace} (see Figure~\ref{fig:inset}).

The design panel is {\textsc{Pail}\xspace}'s primary differentiating feature offering a consistent interface to four design aids, represented as four subsections to the panel, shown in Figure~\ref{fig:inset}:
\begin{enumerate}
    \item \textbf{\dqSec Design Questions \& Goals}: this section tracks the kinds of questions an interaction designer would ask when designing a new interactive piece, including problem formulations, 
    \item \textbf{\crSec Confirmed Requirements}: this section tracks decisions that the human has made or confirmed.
    \item \textbf{\idSec Implicit Decisions}: this section identifies and surfaces decisions the LLM has implemented in the code without explicit confirmation, such as choices regarding how data is structured, or how displayed values are calculated.
    \item \textbf{\uaSec Useful Abstractions}: this section offers grounding terminology, alongside one-sentence descriptions, for the user and the AI to validate that they are referring to the same concepts when they use project-related language.
\end{enumerate}

Each of these subsections contains a list of items generated by the {\textsc{ReflectionsAgent}\xspace} in response to the ongoing conversation in the chat panel, described below. Each item in the design panel includes a justification (``rationale'') for that item as well as 2-3 possible ``alternatives.'' Any alternatives, except some that the {\textsc{ReflectionsAgent}\xspace} identifies as important, are hidden behind an accordion-style view, revealed only when the enclosing item is clicked.

In one study participant's project, within the \dqSec Design Questions \& Goals section, appears the item ``What themes or characters would be most engaging for a 6-year-old?'' accompanied by the rationale ``Engaging themes and characters can make learning more enjoyable and effective.'' and the alternatives ``Animals'', ``Superheroes'', and ``Fantasy worlds with magical creatures''. Finally, each of these alternatives, in turn, is then speculatively assessed by the {\textsc{DesignAgent}\xspace}, which populates each alternative with a sentence about possible cost/benefit trade-offs of that choice. For ``Fantasy worlds with magical creatures,'' for example, the trade-off text is ``Stimulates imagination; may distract from educational content.'' A \tryButton button next to the alternative triggers the {\textsc{DesignAgent}\xspace} to go ahead and make that change. A screenshot of the four sections of the design panel appears in Figure~\ref{fig:design-goals-snapshot}. (An additional \revertButton button appears in response to clicking \tryButton, which restores the original code in the IDE; not shown.)

The design panel is located as the rightmost of {\textsc{Pail}\xspace}'s four main panels (see Figure~\ref{fig:inset}): a \textit{code} panel, an \textit{output} panel (with a console for errors and printed output), a \textit{chat} panel, and the \textit{design} panel. The code panel allows users to view and directly modify any project code, and uses a common in-browser code editor, Monaco,\footnote{\url{https://microsoft.github.io/monaco-editor/}} containing \texttt{p5.js}\footnote{\url{https://p5js.org}} code that is then run and displayed in the output panel.

The chat panel is the interface to a prompted GPT-4o-based chatbot ({\textsc{ConversationAgent}\xspace}) that can read the project code, patch it, or entirely replace it. Changes to code are summarized within the chat, displayed in ``diff'' form, and then propagated directly to the code panel, which shows a holistic ``diff'' over the prior iteration of the project. The {\textsc{ConversationAgent}\xspace} itself is prompted to explicitly guide users through a focused User-Centered Design~\cite{noauthor_what_2024} process: identifying target users, evaluating their needs, assessing possible goals for the project given users' communicated design ideas, and finally generating code for prototypes to test any hypotheses generated through this conversational design process.

The design panel and chat panel operate in a tightly integrated way: conversations in the chat trigger the {\textsc{ReflectionsAgent}\xspace} to update the design panel's contents, while manipulations in the design panel (i.e., trying a specific alternative) trigger code changes via speculative (i.e., uncommitted) executions through the {\textsc{ConversationAgent}\xspace} by the {\textsc{DesignAgent}\xspace}.

\subsubsection{System Design Non-Goals}

Equally important to the design of {\textsc{Pail}\xspace} is what we did \textit{not} include. Because our focus is on design process support, {\textsc{Pail}\xspace} does not provide assistance with debugging or handling nonfunctional LLM-synthesized code, nor any kind of automated QA or simulated user testing, two areas worth considering for an AI-assisted IDE. We leave investigation of these topics to future work. 
\subsection{{\textsc{Pail}\xspace} Implementation} \label{sec:implementation}

\aptLtoX[graphic=no,type=html]{\begin{figure*}
    \includegraphics[width=0.8\textwidth]{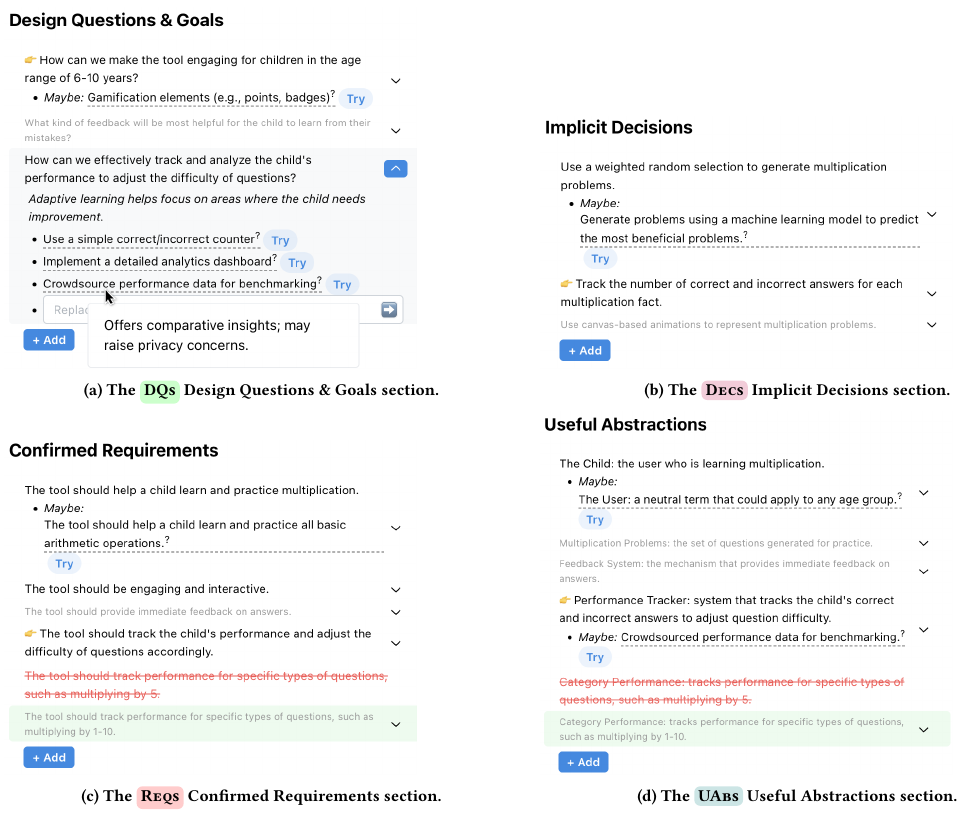}
    \caption{A screenshot of the contents of each of the four components of the design panel, from P9. The third question in \dqSec \textit{Design Questions \& Goals} \textbf{(a)} (``How can we effectively [...]?'') is expanded to show the rationale for that question along with three alternatives. The mouse hover over the third alternative (``Crowdsource [...]'') reveals a speculated trade-off for trying that option.  In {\textsc{Pail}\xspace}, these four sections are organized vertically in the design panel. Items with a finger are \textit{important}, dotted underlines allow hover (see \textbf{(a)}) for trade-offs. Green background and red strikethrough indicate recent changes.}
  \Description{A snapshot of the design panel. Some alternatives are identified by the {\textsc{ReflectionsAgent}\xspace} as important; these are shown under the ``Maybe:'' heading, inline without expansion.}
  \label{fig:design-goals-snapshot}
\end{figure*}}{\begin{figure*}
 \begin{subfigure}{0.49\textwidth}
     \includegraphics[width=0.8\textwidth]{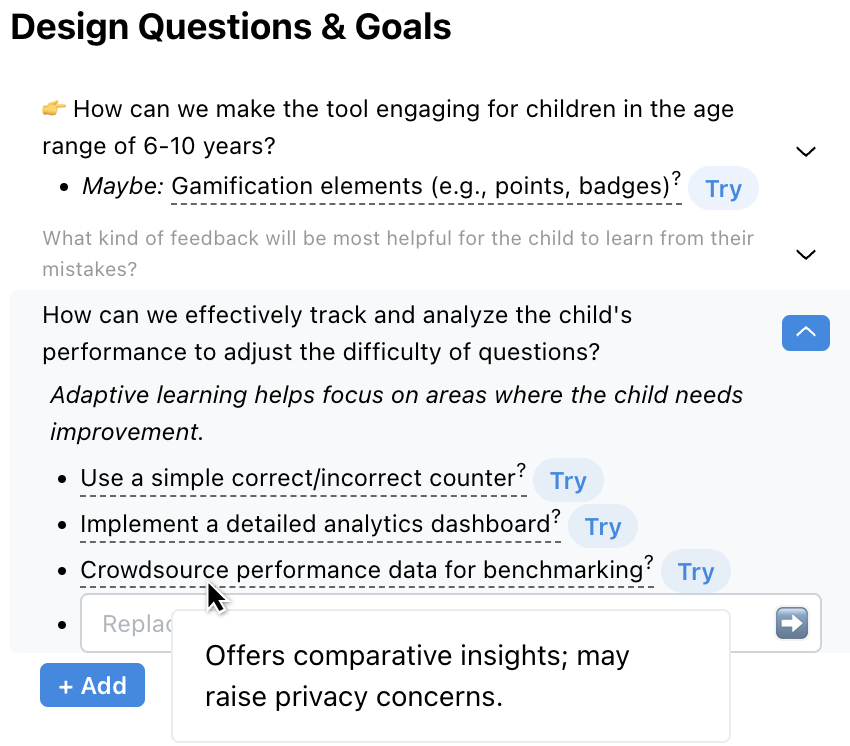}
     \caption{The \dqSec \textbf{Design Questions \& Goals} section.}
     \label{fig:a}
 \end{subfigure}
 \hfill
 \begin{subfigure}{0.49\textwidth}
     \includegraphics[width=0.8\textwidth]{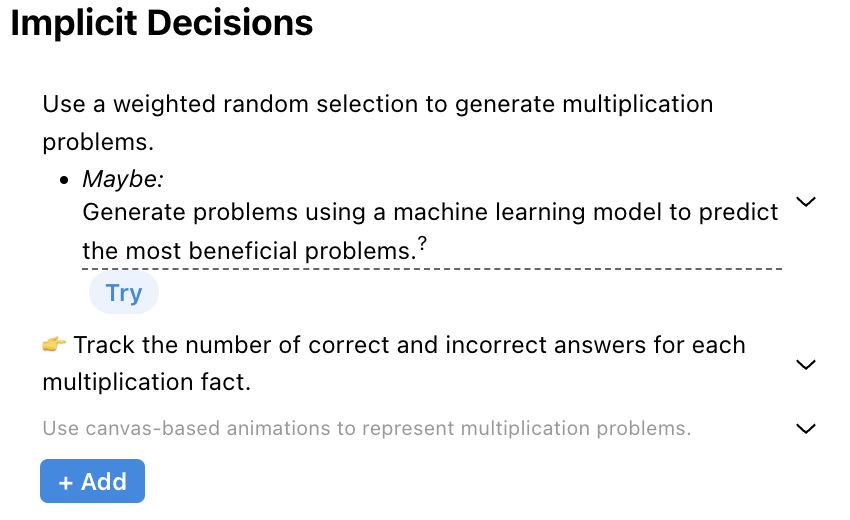}
     \caption{The \idSec \textbf{Implicit Decisions} section.}
     \label{fig:c}
 \end{subfigure}
 
 \medskip
 \begin{subfigure}{0.49\textwidth}
     \includegraphics[width=0.8\textwidth]{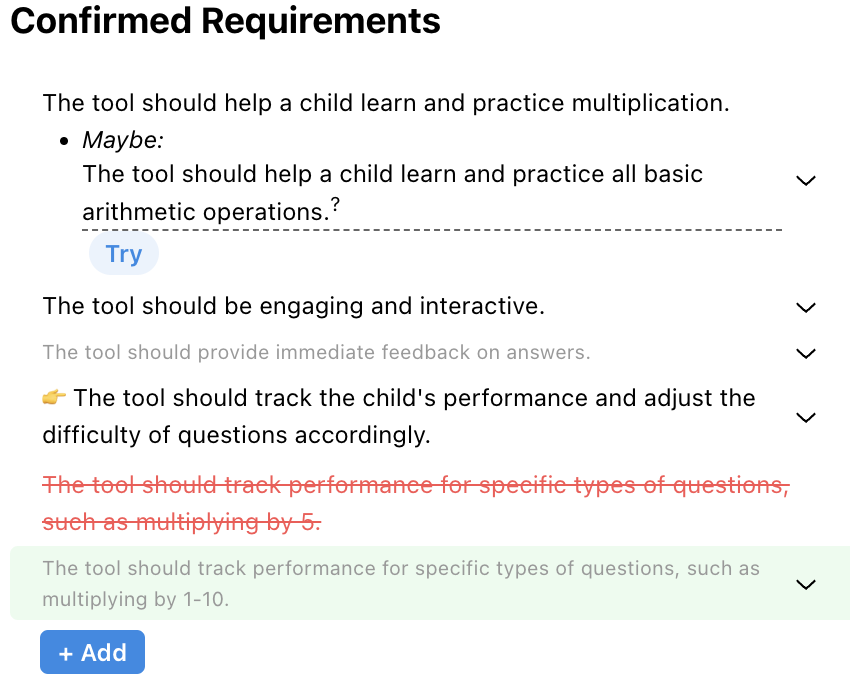}
     \caption{The \crSec \textbf{Confirmed Requirements} section.}
     \label{fig:b}
 \end{subfigure}
 \hfill
 \begin{subfigure}{0.49\textwidth}
     \includegraphics[width=0.8\textwidth]{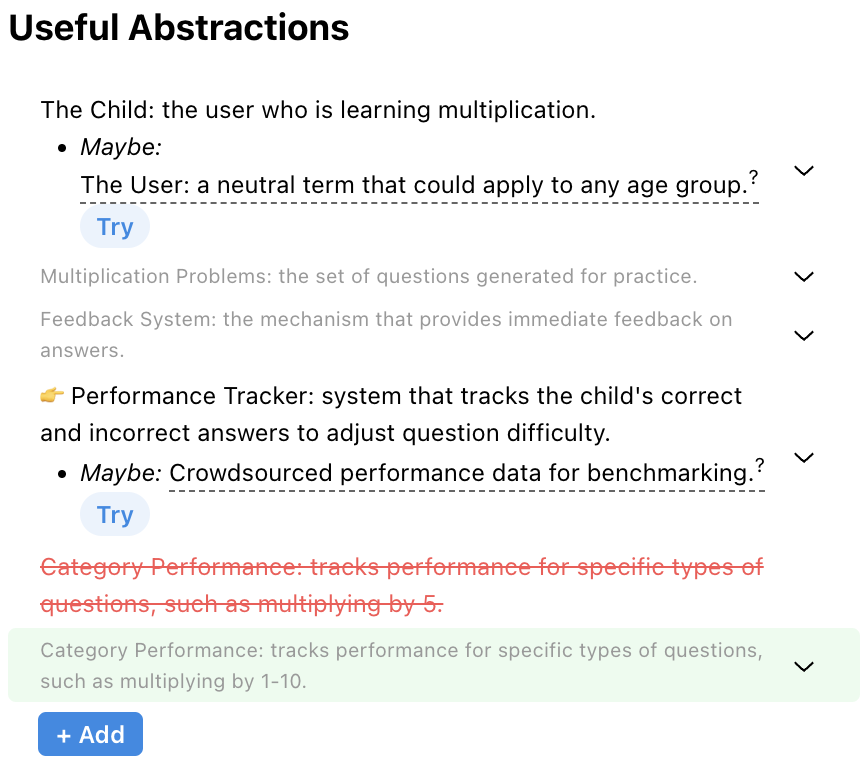}
     \caption{The \uaSec \textbf{Useful Abstractions} section.}
     \label{fig:d}
 \end{subfigure}

    \caption{A screenshot of the contents of each of the four components of the design panel, from P9. The third question in \dqSec \textit{Design Questions \& Goals} \textbf{(a)} (``How can we effectively [...]?'') is expanded to show the rationale for that question along with three alternatives. The mouse hover over the third alternative (``Crowdsource [...]'') reveals a speculated trade-off for trying that option.  In {\textsc{Pail}\xspace}, these four sections are organized vertically in the design panel. Items with a finger are \textit{important}, dotted underlines allow hover (see \textbf{(a)}) for trade-offs. Green background and red strikethrough indicate recent changes.}
  \Description{A snapshot of the design panel. Some alternatives are identified by the {\textsc{ReflectionsAgent}\xspace} as important; these are shown under the ``Maybe:'' heading, inline without expansion.}
  \label{fig:design-goals-snapshot}
\end{figure*}}

{\textsc{Pail}\xspace} is implemented as a React single-page application, proxying calls to GPT-4o through a node.js backend that logs all requests. The code editor uses Microsoft's Monaco\footnote{\url{https://microsoft.github.io/monaco-editor/}} editor configured to display code differences inline. The chat component is a custom-built turn-taking component for user communication with the {\textsc{ConversationAgent}\xspace}; it shows chat messages as well as summaries of any code changes made by the {\textsc{ConversationAgent}\xspace} alongside diffs with any changed code.

We chose the Monaco editor (a component from VS Code) specifically because of its overall flexibility and support for showing ``diffs'' of code inline. This code is run using the \texttt{p5.js} library run alongside an HTML Canvas element enclosde within an \texttt{iframe} element, using the \texttt{loop-protect} library\footnote{\url{https://www.npmjs.com/package/loop-protect}} to prevent infinite loops from interrupting event processing in the browser. The iframe's \texttt{console} object (used to report errors to developers) is proxied through the browser's \texttt{postMessage} API to allows p5.js errors and other console messages to pass through to the IDE.

A few technically challenging parts of {\textsc{Pail}\xspace}'s implementation are worth noting.

\subsubsection{Speculative Assessment}

The code and design panel JSON structures are stored as separate ``artifacts'' in versioned flat files by the node.js server. To improve latency and offer an \textit{a priori} indication of the impact a particular change might make the {\textsc{DesignAgent}\xspace} speculatively assess a subset of identified alternatives to design panel items. 

These speculative assessments vary in scope; depending on the nature of the alternative, this could include: (1) how that alternative might impact \dqSec design goals, such as considerations of target users; (2) what additional \crSec requirements that alternative might reveal; and (3) any new \dqSec design \textit{questions} that may emerge from consideration of that alternative. Figure~\ref{fig:agents} shows how agents and artifacts influence one another.

\subsubsection{Differential Updates to Code}

At the time {\textsc{Pail}\xspace} was created, LLMs excelled at generating full files worth of code, but did not have a mechanism for incrementally updating existing code. Because {\textsc{Pail}\xspace} is premised on rapid iteration across levels of abstraction, we decided it was important to have such a mechanism, so that users could ask for a small tweak (e.g., ``can we make the small rectangles blue?'') and have that change implemented quickly. Other systems like Claude's \textit{Artifacts} and ChatGPT's \textit{Canvas} rewrite the full code from scratch each time, which is not only slow, but also often includes other, unrelated changes to the code---a complaint we also heard in our formative study.

Obvious solutions we tried, like asking an LLM to output a code change in \textit{diff} or \textit{unified diff} format, failed frequently: line numbers wouldn't match up, edits were interleaved with both old and new code, and new code was sometimes inserted in the wrong place altogether. These errors align with some prior observations of LLMs' failure modes too: arithmetic is not a strong suit; text locality matters and disrupting it (by interleaving new and old code, for example) lowers performance.

{\textsc{Pail}\xspace} instead provides all active code to the LLM with every request, transforming that code by prefixing \texttt{L\#\#\#\#} to each line, with \texttt{\#\#\#\#} replaced by each line's consecutive line number---thus providing a built-in line-numbering scheme. Edits, implemented as a tool call, are represented as ``old code'' using the same \texttt{L\#\#\#\#} prefix, and ``new code'', again with a line number prefix. This allows a straightforward extraction of which lines to replace (fuzzy-matched against line numbers and old code lines, for robustness), and locality in the production of new code. With this technique, we found that {\textsc{Pail}\xspace} succeeds in patching code directly from LLM calls over 90\% of the time, with failures readily discoverable---and full-rewrite as a backup. As a bonus, these incremental updates can be streamed to users as well.

\subsubsection{Differential Updates to Design Artifacts}

In addition to incremental changes to the code, we implemented incremental changes to the four components of the Design Panel. Here, the challenge was primarily one of updating the visualization: these components are regenerated from scratch often, and in phases (first the summary line, then the alternatives, rationales, and highlights)---as with updates to code, we wish to begin displaying this data to users immediately as it comes in, especially when there is not already any data to display.

But when there is already-displayed data to update, {\textsc{Pail}\xspace} needs to evaluate the new set of, e.g., \crSec Requirements, which are streaming in as an incomplete JSON object, against a full set of existing \crSec Requirements complete with alternatives, rationales, and highlights. This is accomplished by fuzzy progressive prefix-matching of incoming data with existing data. For \crSec, for example, individual requirements coming in through the stream are compared with existing requirements for equivalency up to a threshold; only when confidence is high that a requirement is new does the visualization update to display the new requirement as its contents stream in from the {\textsc{ReflectionsAgent}\xspace}. Meanwhile, prefix-matched requirements (such as an incoming requirement with the incomplete text ``Track the num'' matching ``Track the number of correct and incorrect answers.'') are not updated until changes are detected, avoiding re-rendering of existing requirements as they are streamed in. This technique is then applied, recursively, to the nested alternatives, rationales, and highlights.


\section{{\textsc{Pail}\xspace} Usage: An Example}

To make the design and intended use of {\textsc{Pail}\xspace} concrete, consider the following usage scenario. Sam, an experienced backend system software engineer and parent of two---kids 3 and 5 years of age---is eager to help her 5-year-old, Alex, learn how to read simple words. Sam's tried what feels like all the iPad apps in the store, but none have clicked with Alex. Sam knows she could write a new app, but she's not sure she can commit the time: programming in an unfamiliar environment (e.g., a game environment) often has a learning curve with high variance in time required, and there would likely be a lot of upfront preparation required before Sam could show Alex anything actually interactive---time ultimately wasted if Sam's game ideas don't appeal to Alex.

With {\textsc{Pail}\xspace}, Sam engages in the following scenario:

\begin{enumerate}
    \item Sam asks the {\textsc{ConversationAgent}\xspace} for help making a game for Alex to practice reading simple words.
    \item {\textsc{Pail}\xspace} responds with a chat message containing a few questions; a few seconds later, these appear in the \dqSec Design Questions section of the design panel.
    \item Sam scans the suggestions in the \dqSec list: about who the user is, what kind of help they need, and what her own goals are for the app, finding the following:
    \begin{itemize}
        \item[$\bullet$] \textit{What specific skills should the game focus on? (E.g., letter recognition, phonics, simple words)}---with an explicit suggestion she try \textit{Focus on simple words}.
        \item [$\bullet$]\textit{What types of activities or game mechanics would be most engaging?}---with the note that engagement will keep the child motivated.
        \item [$\bullet$]\textit{Are there any themes or characters that your child particularly likes that we could incorporate?}
    \end{itemize} See Figure~\ref{fig:design-goals-snapshot} for a screenshot of a similar set of \dqSec Design Goals.
    \item Sam considers what she knows about Alex, what kind of support she needs, and more. Ultimately she clicks \tryButton next to the \dqSec Design Question alternative \textit{Focus on simple words.}
    \item {\textsc{Pail}\xspace}'s {\textsc{DesignAgent}\xspace} notes this suggestion and follows up with another question: \textit{Any kinds of activities your kid enjoys?}
    \item This time, Sam responds in chat: \textit{maybe a word-picture matching game?}
    \item {\textsc{Pail}\xspace}'s {\textsc{ConversationAgent}\xspace} offers an idea for an implementation using a drag-and-drop interaction, asking \textit{does this approach sound good to you?}
    \item Sam finds this suggestion confusing. Drag-and-drop isn't an interaction Alex is that familiar with, and in other apps it hasn't worked very well. What are some alternatives? Sam scans the design panel and finds drag-and-drop interaction in the \idSec Implicit Decisions section of the design panel. She clicks \tryButton next to \textit{Use a tap-to-select interface for matching words to pictures.}
    \item {\textsc{Pail}\xspace}'s {\textsc{DesignAgent}\xspace} has speculatively assessed this approach already, and identified that tap-to-select is \textit{simpler for very young children, and can be more intuitive than drag-and-drop}---noting this when Sam was scanning through possible alternatives.
    \item {\textsc{Pail}\xspace} moves this item up to \crSec Confirmed Requirements and produces an initial prototype of the matching interaction: simple words and simple emojis, in a single line. The {\textsc{ConversationAgent}\xspace} offers a summary of code changes, and the {\textsc{ReflectionsAgent}\xspace} records that \idSec \textit{Words and pictures are displayed in a simplified layout.}
    \item Sam doesn't love this choice, and clicks \tryButton next to the \idSec alternative \textit{Display words and pictures in a grid layout}.
    \item In response to {\textsc{ConversationAgent}\xspace}'s generated patch to use a grid layout, {\textsc{Pail}\xspace}'s {\textsc{ReflectionsAgent}\xspace} also moves this new requirement to the \crSec Confirmed Requirements section.
    \item At this point, Sam shows the prototype to Alex to gather feedback: are the pictures understandable? Are the words too complex, or too simple? 
    \item Sam and Alex can then work together to revise the project, incorporating Alex's feedback about what he likes and doesn't like about using the game.
\end{enumerate}

Note that nothing here is beyond Sam's capabilities as a software engineer. She knows how to write code, but is unlikely to find typical ``paper prototyping'' worthwhile for this use case. {\textsc{Pail}\xspace} gives Sam the ability to ``sketch'' with code, overcoming the activation energy required to choose and initialize an environment, figure out a data model and rendering pipeline, or otherwise make a large number of \textit{boilerplate decisions} that must be resolved to have a working program, but serve no other purpose towards Sam's goals around understanding what will help Alex learn to read.

Instead, using {\textsc{Pail}\xspace}, in 10 minutes Sam has already learned that Alex adores emoji iconography but the targets are too small for his fingers. Sam and Alex can continue using {\textsc{Pail}\xspace} to iterate on the game, pursuing completely new directions without losing track of the lessons they learned from earlier prototypes. {\textsc{Pail}\xspace}'s suggested alternatives enable Sam and Alex to break free of anchoring effects or the ``sunk code'' fallacy of energy invested in authoring code. Using {\textsc{Pail}\xspace}, the energy invested is in considering alternatives, exploring the design space with potential users, and testing prototypes---epistemic actions resolving unknowns, which serve a useful design purpose whether they resolve positively or negatively.


\section{User Study Procedure}
We had two goals in our user study: 
(1) identify possible ways the various built affordances helped or did not help users (though not prove definitively); and 
(2) identify likely challenges inherent to \edits{AI support for design.}

\subsection{Interview Protocol}

We began each interview by showing a demo use case for {\textsc{Pail}\xspace}, with an emphasis on what we expect it to be helpful for: designing interactive applications. We showed how to use the chat functionality, explained the tight integration between the chat agent and the code (as mentioned in \S\ref{sec:implementation}, the chat agent can directly modify the code), and then walked users through the various parts of the design panel, explaining each subsection. With each subsection, we showed examples of entries, complete with the rationales, lists of alternatives, and results of speculative execution, and, finally, demonstrated what happened when the user clicks on the \tryButton button. We finished our introduction by explaining our research question, answering any questions from participants, and then moving on to the first task.

We encouraged participants to think of {\textsc{Pail}\xspace} not only as a \textit{code generation} tool, but also as a \textit{design} tool. We informed participants that they could ask for broad, high-level goals in the chat panel, such as ``Help me design an interactive feature that goes along with this article: [pasted article]'' (see \S\ref{task:article}, Task 1 below)---that the chatbot was designed to walk them through a design process, considering target users and the participant's own high-level goals.

Because our primary goal is not to measure the effectiveness of {\textsc{Pail}\xspace} from a typical ``HCI system evaluation'' perspective, but rather to learn what the point design that {\textsc{Pail}\xspace} represents can tell us about the broader design space of AI-assisted programming, we actively encouraged participants to make use of {\textsc{Pail}\xspace}'s affordances, and gave them suggestions on when and how to do so. This allowed us to observe a broader range of interactions and impacts than a more traditional lab study, or which might otherwise only be visible after participants develop an expertise using {\textsc{Pail}\xspace}. Naturally, this choice comes with limitations on what we can thus claim.

\subsection{Tasks}

Our primary criteria for task selection were: ambitious and ambiguous goals; robustness to common LLM failure modes, such as tracking long (or multiple) code files; and enough design focus that we could reasonably observe a diversity of design approaches among our participants.

As mentioned in \S\ref{sec:designGoals}, because we are primarily interested in the impact of design support on program design, and not on LLMs' code synthesis capabilities or ability to support end users in learning programming concepts, we chose tasks that were ambitious but achievable using a length of code that our choice of LLM could robustly handle. These interests also drove selection of our participant pool, discussed below.

We ultimately selected 3 tasks for our participants:

\begin{enumerate}
    \item[Task 1:] Create an interactive feature to go along with an article about the impacts of air conditioning on migration patterns in high-average-temperature areas. \label{task:article}
    \item[Task 2:] Create a game that helps a young child learn how to read or multiply.
    \item[Task 3:] Create a simulation to show the effects of medical overtesting, e.g., recommending screening for specific rare diseases for more people.    
\end{enumerate}

For participants who were parents, we started with task 2, and specifically asked for them to consider the needs of their own child and create a game to address one such need, so that we could observe whether having a very specific target user, who is well-known to the designer, has an impact on the design process. Parents were then asked to engage in task 1, if time permitted. All other participants were asked to engage in Task 1 first, then offered a choice of 2 or 3.

Several participants also requested time to engage in an open-ended exploratory task, in which they could experiment with {\textsc{Pail}\xspace} in service of some personally meaningful goal; we supported this where time allowed. 
\subsection{Participants}
Participants were a mix of 5 academics and 6 professionals, of whom 3 were parents. We recruited participants with varying levels of programming expertise, design experience, and prior LLM use; see Table~\ref{tab:participants}. Our sample size was chosen in line with prior work around formative testing for usability~\cite{faulkner_beyond_2003,sauro_quantifying_2016}: our goal is to explore the early benefits and likely challenges users encounter when engaged in our task using {\textsc{Pail}\xspace}, and our experience with pilot users suggested that we would find benefits and challenges quite quickly.

Our participant pool is skewed towards, but not exclusively consisting of, professionals and graduate students in design- and STEM-related fields---and it is certainly not representative of the population at large. We are not making claims about how a specific population does or does not engage in specific behavior, but rather seek to identify {\textsc{Pail}\xspace} benefits and forecast future challenges among a population that we expect is disproportionately likely to be early adopters of LLM-based tools.

\edits{We also skewed our pool towards participants with programming experience because we are most interested in design support, and wanted to avoid confounding factors caused by participant inexperience with code, programming systems, or other parts of {\textsc{Pail}\xspace} that were not directly related to our research goals.}


\begin{table*}[t]
\centering
\renewcommand{\arraystretch}{1.2} 
\begin{tabular}{c|c|c|c|c|c|c} 
\toprule
\textbf{ID} & \textbf{Age} & \textbf{Parent?} & \textbf{Professional Background} & \textbf{Design Experience} & \textbf{Programming Experience} & \textbf{LLM Use}
  \\ 
\hline
P1 & 40s & Yes & Design Researcher & Professional & Professional & Infrequent \\
P2 & 20s & - & HCI Graduate Student & Professional & Professional & Frequent \\
P3 & 30s & - & HCI Researcher & Amateur & Professional & Frequent \\
P4 & 30s & - & Software Eng Manager & None & Professional & Infrequent  \\
P5 & 50s & - & Software Eng Manager & Professional & Professional & Moderate \\
P6 & 60s & - & Film \& Media Artist & Professional & Amateur & Frequent \\
P7 & 20s & - & HCI Graduate Student & Professional & Amateur & Infrequent \\
P8 & 40s & Yes & Filmmaker \& Academic & Limited & Limited & Infrequent \\
P9 & 50s & - & Sound Artist \& Lecturer & Professional & Amateur & Infrequent \\
P10 & 60s & - & Software Engineer & Amateur & Professional & Moderate \\
P11 & 50s & Yes & Software Engineer & Limited & Professional & Moderate \\
\bottomrule

\end{tabular}
\captionsetup{width=0.95\linewidth}
\vspace{0.2cm}
    \caption{Study participants.}
    \label{tab:participants}
\end{table*}


\subsection{Analysis \& Evaluation}

Participants were instructed to think aloud while engaged in the tasks. We then undertook an exploratory data analysis, transcribing all videos and observing where participants directed their attention, where they made design decisions, and what factors appeared to influence them. We also noted where interviewers intervened. We then compared these observations across participants and categorized their approaches and uses of {\textsc{Pail}\xspace} using a well-established affinity diagramming process~\cite{moggridge_designing_2006}, and through the use of service blueprints~\cite{bitner_service_2008} that document participants' behavior and reflections. We elected to use these methods from HCI and service design because we seek to understand broadly how different participants engage in similar tasks with a new technology---one that does not have an existing set of users with pre-established work practices they are already engaged in.

\section{Findings}

In this section, we report findings from our participants' use of {\textsc{Pail}\xspace}. In particular, we find that \edits{(1) participants frequently engage in rapid-fire repeated iteration, commonly at higher level of abstraction than code, and often (but not always) through use of {\textsc{Pail}\xspace}'s Design Panel. While in this state, participants treated iterations as disposable ``sketches'', demonstrating little attachment, and engaged in activities spanning all parts of the ``4 D's'' of design, from problem discovery and definition through solution development and delivery, sometimes within the same action.} 

We also find that (2) focused attention was spread quite thinly while participants used {\textsc{Pail}\xspace}, and between changes to code, new messages from chat, and changes to the design panel contents, awareness of what was happening in {\textsc{Pail}\xspace} was easily lost and was perceived as costly to regain---and that these perceived costs ultimately influenced where participants directed their attention.

\edits{In \S\ref{sec:pail-use}, we describe how participants use (or do not use) {\textsc{Pail}\xspace}'s design support to move across abstractions, consider alternatives, and explore the spaces of possible problems and user needs. Then, in \S\ref{sec:pail-stumbles} we examine where {\textsc{Pail}\xspace} falls short in supporting users' design processes, and why.}

\subsection{Use of {\textsc{Pail}\xspace}'s Design Support} \label{sec:pail-use}

First, we report on ways in which participants used {\textsc{Pail}\xspace}, drawn from in-interview reflections and post-session analysis of {\textsc{Pail}\xspace} use.
Across our open-ended program design tasks, participants were quite varied in their approaches. 

A common (9/11) initial stumbling block was choosing the first action: though all participants were shown a ``project in progress'' within {\textsc{Pail}\xspace} with an overview of its various components (see Fig.~\ref{fig:inset}), most (7/11) were not sure at what level of abstraction to make their first request. Should they think of a solution first, and then request that solution? Should they simply provide a description of the design task (and in the case of article-associated interactive task, the article) directly to the {\textsc{ConversationAgent}\xspace}? Only four participants began by asking the {\textsc{ConversationAgent}\xspace} about the design task directly, even though we consistently provided guidance that {\textsc{Pail}\xspace} could help them brainstorm ideas, too, and that they should not feel the need to wait until they had a concrete idea to start making requests.

\subsubsection{Abstracting Up \& Problem Exploration}

Regardless of the nature of the initial request, the {\textsc{ConversationAgent}\xspace} would then respond with questions aligned with the design process described above, asking about \dqSec design goals, target users and user needs, desired outcomes, and often offering some plausible responses for each. For some participants, this would be the first time they would step back to reckon with these factors explicitly; even participants who had thought about what specifically to design often did so without discussing goals or users, but rather brainstorming solutions directly based on what each solution could provide.

All participants interacted with the initial ``design process'' phase of the {\textsc{ConversationAgent}\xspace}, and all were steered, to varying degrees, by being prompted with these questions. Participants who started by suggesting a specific point design would often use these questions as an opportunity to think more broadly. Even senior programmers with design expertise found something to consider in {\textsc{ConversationAgent}\xspace}'s prompting, expressing thoughts like ``this really broke it down in an interesting way,  who is your target audience'' (P5). Later, P5 would reflect:
\begin{quote}
    I didn't have an idea of what I wanted to do when you first presented me with this problem, and so putting this in and then kind of exploring the "oh, i see, it came up with regions"---and I thought "that sounds reasonable let's start with regions.
\end{quote}

The questions could be an awkward fit for some participant-initiated tasks, however. For example, in tasks that were not targeted at user needs, the questions seemed not well-targeted to the participant's goals. P6 wanted to use {\textsc{Pail}\xspace} to make a particular ``artistic'' sketch they had in mind using \texttt{p5.js}, with which they were already familiar. Recalling their reckoning with the questions {\textsc{ConversationAgent}\xspace} posed, P6 later reflected:
\begin{quote}
    Yeah I guess the design element, meaning the ``design'' as the [...] medium that you're working with here, and having these different questions, requirements, decisions, useful abstractions and things like that, is pretty interesting \textit{[long pause]} But I like it, I like being turned on my ear, you know, it's good.
\end{quote}

Some participants also wanted to start directly with an example, rather than by thinking through user groups and user needs. One participant, before even using {\textsc{Pail}\xspace}, rationalized this desire by noting that it was easier to iterate from a single design than to come up with one from scratch.

Typically, participants would then consider these questions, responding either directly to the {\textsc{ConversationAgent}\xspace}, or scanning the summary of questions and possible answers in the design panel and then exploring a subset of {\textsc{ReflectionsAgent}\xspace}-proposed alternatives through the {\textsc{DesignAgent}\xspace}. Participants varied in how they responded to these questions when they did engage. For example, P11 treated the design questions and proposed answers as a checklist to select user groups and features from, clicking the \tryButton button on all the options that appealed, upfront, and only then ran the generated program.

\paragraph{Reflection} Recall that one of our major goals with {\textsc{Pail}\xspace} was to encourage thinking about design goals and questions explicitly---on this count, we succeeded for many participants, but not all. It is clear from our study that some users are likely to want to start directly from a point solution, and only then reconsider design goals, target users, etc.---and future tools should consider how to best serve this population, perhaps by starting from a set of easily-comparable point solutions to reduce anchoring.

\subsubsection{A Working Sketch: First Contact \& Rapid Iteration}

For most participants, the first prototype that instantiates a solution, for a sufficiently-formulated problem statement, is revelatory---exposing a number of mismatches between the participant's understanding of the project and either the current state of the code, or the agents' understanding of the project. For example, P9 expected a set of question and answer cards to be shuffled evenly across the canvas, but found them separated by category into distinct question and answer regions. Two other participants (P5, P7) converged on map views with the {\textsc{ConversationAgent}\xspace}, but then saw initial prototypes that didn't include maps \textit{per se}, but rather stylized region diagrams with rectangles and triangles representing world or country regions.

These types of mismatches almost always resulted in a flurry of requests for low-level implementation fixes. Participants typically requested these fixes either directly in natural language from the {\textsc{ConversationAgent}\xspace} or by finding a suitable alternative in the design panel and trying it.
How long participants spent in this rapid-fire local iteration mode varied, from under a minute (P8) to more than 20 minutes (P7), during which their activities resembled a \textit{flow state}~\cite{czikszentmihalyi_flow_1990}. These participants appeared to have deep concentration, could express what they wanted as next steps rapidly, noted an effortlessness to the repeated iteration, and with minimal rumination. Participants' think-aloud would often pause in this state. \label{sec:flow}

In this state, participants would often rapidly shift attention across the code, output, chat, and design panel, looking to make sense of what they were seeing, and for how to communicate desired changes most effectively. Those with greater programming expertise, or whose expertise was a closer match with the task domain, unsurprisingly spent more time looking at the code, but not more time editing it directly---rather, seeking confirmation, in the code, of a ``lack of surprises'' (P3) to validate that the program was being authored in a way that met participants' expectations.
\edits{Meanwhile, t}hose with less domain expertise (e.g., no familiarity with p5.js, or limited familiarity with authoring interactive artifacts) often ignored the code entirely past a certain point (7/11 participants). 

Because none of the agents were designed (nor inclined) to break users out of this state, it would often continue until either some insurmountable ``blocker'' would interrupt, such as a bug the {\textsc{ConversationAgent}\xspace} couldn't fix, or the user reached a point where they achieved whatever prototype they had set out to achieve, and needed to reflect on what to do next.
However interrupted, participants would next take stock of where they were, deciding whether to continue making iterative changes, or reconsider the current approach's suitability towards higher-level goals. 

\paragraph{Reflection} This rapid iteration was almost always quite broad, and not limited to any one of the traditional ``4 D's'' of the Double Diamond: the \textit{discover, define, develop,} and \textit{deliver} phases of design. In fact, the same action might serve multiple goals: validating the defined problem while simultaneously progressing towards prototype development, for example, or revealing some new, unanticipated end-user need. To the extent that we expected to support design processes, it appears that rapid iteration with \textit{interactive prototypes} \edits{as sketches enables a metaphorical \textit{superposition} of the problem-formulation and solution-exploration phases of design, perhaps enabling more rapid iteration than paper sketches or other traditional \textit{discovery} methods in design. We are not suggesting that these traditional methods of sketching and prototyping are obsolete, but rather noting that these interactive prototypes were being used in a manner similar to an ink-on-paper sketch in an architectural design setting: as disposable, low-investment \textit{sketches} that allow the designer to investigate one facet of a design space without needing to simultaneously resolve decisions across the full design space.}

\subsubsection{The Design Panel, Alternatives, and Rationales}

Nearly all (10/11) participants reported finding the design panel useful. Eight participants appreciated the summarization and tracking affordances that enabled quick scans of the (\dqSec,\crSec) decisions made so far (and updates to those decisions) when new chat messages began to exceed in length how much participants desired to read. Seven participants expressed appreciation for the reporting of rationales and alternatives. Though no participants explicitly directed appreciation towards the \idSec implicit decisions or \uaSec useful abstractions components of the design panel, we nonetheless observed several participants (P3, P5, P8, P9, P11) drawing insight from these components. P2, for example, explicitly drew a comparison with their prior ChatGPT experience:
\begin{quote}
    I did try to use ChatGPT for that and it was...fine. [...] It looked [worse] than the one we just built, and it took me longer. [...] [In ChatGPT] I didn't have the right mechanism to high-level changes at this level of abstraction.
\end{quote}

Scanning the alternatives lists (and ultimately selecting an alternative to \tryButton) typically emerged after some initial trigger, as when a participant would pause, unsure of what next step to take---in a broader sense than just making the next single decision. For example, at one point P10 recognized that their approach to visualizing one particular set of interrelated values (in their case, temperature and air conditioning usage) wasn't going to work, and they weren't sure where to take the project next. In this and other similar cases, the decisions and lists of alternatives offered a lower-cognitive-load path forward, by allowing participants to \textit{select one of several possible options} rather than \textit{generate a new idea entirely} as a next step, echoing the recognition-over-recall UX design heuristic~\cite{nielsen_10_nodate}. Scanning these lists would often result in focus on a single choice, eliciting a reaction like ``oh that seems like a good idea'' (P4)---or would yield an alternative not directly on the lists, but still inspired by the scan, which participants would then suggest either in the chat or in the alternative lists' open text field titled ``Replace with...''.

A few participants (P5, P7, P11), in ``flow state'' (see \S\ref{sec:flow}), preferred to use the design panel's alternatives almost exclusively to guide their exploration, avoiding chat. Asked why, P5 reported that it was easier to answer multiple-choice questions than to repeatedly write messages in chat---treating the lists of alternatives in the \dqSec design questions section as a sort of ``design checklist,'' a menu from which to select target users, goals, and more. 
\paragraph{Reflection} To the extent that we expected the design panel to encourage participants to consider alternatives \textit{prospectively}, our findings here and in the previous subsection suggest we could do better: participants only rarely actively sought out new problem formulations unless prompted either by {\textsc{Pail}\xspace}, by an interviewer, or by a realization that their current design approach was not going to lead to success. \edits{Instead, they appear to consider alternatives only \textit{retrospectively}, after exhausting a particular, though narrow, line of inquiry.}

\subsection{Stumbles, Mismatches in Participants' Design Processes} \label{sec:pail-stumbles}

Beyond participants' use of {\textsc{Pail}\xspace}'s specific affordances, we also observed behavior that bears on the design of future AI assistance for program design; here, we detail those observations.

\subsubsection{Re-considering Design \textbf{Problems}}

Though {\textsc{Pail}\xspace} explicitly elicits design considerations from participants at the outset, there is no explicit support to bring users to \textit{reconsider} design questions and goals during the implementation process. As a result, few participants explicitly reconsidered the highest-level design directions within {\textsc{Pail}\xspace}, at least without explicit prompting from an interviewer. In fact, in the ``flow state,'' many participants would fixate and iterate on small details (e.g., colors, item position, text content) repeatedly. Meanwhile, the {\textsc{ConversationAgent}\xspace} was happy to support this low-level iteration for at least as long as interviewers allowed. 

But it was \textit{not the case} that participants remained fixed in their beliefs about user needs and which is the right problem to solve---they simply did not reconsider their design goals \textit{explicitly in conversation} with the {\textsc{ConversationAgent}\xspace}. Instead, these realizations would come from specific prototypes that yielded specific forms of insight, such as whether particular problem formulation could be compellingly addressed. For example, P10 at one point reflected: can an end-user actually be convinced of a causal relationship between air conditioning and climate change through a bar chart---or should the interactive feature instead focus on a different climate-related relationship?

For most participants, this reconsideration was more often voiced to the interviewer rather than to the {\textsc{ConversationAgent}\xspace}, highlighting the need for explicit elicitation, at least initially.

\subsubsection{Content Generation and Overwhelming Information}

Nearly every participant at some point commented on the overload generated by {\textsc{Pail}\xspace}. 
There is too much data, it is being generated too fast, it's too hard to look at everything, and it's not clear what one should be looking at. As a result, participants reported, much effort was spent figuring out where to look and how to evaluate changes. Running the project was almost always the top choice, but that did not always work. Sometimes, program behavior was not trivial to reproduce; bugs prevented visual output; or the program simply could not be run because the {\textsc{ConversationAgent}\xspace} was in the process of updating it, which could take longer than a minute for large changes. While waiting, participants often scanned the design panel, or the ``diff'' view of the code, to understand the scope of recent changes or consider what steps to take after the code updates have completed. 
In P5's words: ``Like, it's asking me so much in here, I'm not gonna read it every time.''

\edits{The main challenge causing a feeling of overwhelming information between the code, chat, and design panel affordances, participants reported, was simply too much data coming in at once. When participants stopped to read through the design panel contents, they could do so without a sense of overwhelm---and even the chat's relatively low signal-to-noise ratio content could be read in this way. Rather, the challenges arose when clicking a \tryButton button or making a request, which would trigger a cascade of changes in the interface; it was this cascade in particular that participants struggled with.}

\subsubsection{Application of Reflection and Design Agents}
Third, participants did not appear to make substantial use of the rationales the {\textsc{ReflectionsAgent}\xspace} and {\textsc{DesignAgent}\xspace} provided for why certain decisions were made or what trade-offs would result form selecting a particular alternative. When asked, participants reported that they simply found it \textit{easier to try an alternative} than to consider whether the provided rationale was valid and relevant---a version of ``show, don't tell.''~\cite{swan_how_1927}

\edits{Participants also reported feeling little reason to \textit{trust} these rationales, which were restricted in how much detail they provided. This lack of detail limited the expected epistemic value of the proclaimed alternative compared with manual testing, but it also made it challenging for participants to understand \textit{on what basis} those rationales were generated, a critical input to participants' assessment of validity.}

\subsubsection{Attention, Expertise, and the Cost of Awareness}

As the contents of the code, chat, and design panes updated ``automatically'' through agent updates, staying on top of the latest updates to any particular pane required substantial attention. Once lapsed, this ``awareness'' was costly to regain. The exact cost depended on expertise, all else being equal. For example, senior programmers rapidly lost awareness of the code, and their expertise helped them regain it quickly when needed. The cost of regaining awareness depended on participants' choices for where to direct attention. If the code felt ``hopeless'' (P1), or ``unfamiliar'' (P5), regaining awareness became a priority only when a participant encountered a bug or issue. 
P5 described the experience of clicking \tryButton and watching the code change in response, in the following way:
\begin{quote}
    Each time I click on something here [in the design panel] [...] I'm like ``Ah! What part of this is important?''
\end{quote}

Both domain and programming expertise played a role in mitigating those costs, and thus in how effectively participants could stay on top of changes to code and design. This effect manifested in a few ways: first, domain expertise helped participants more easily recognize the overall ``shape'' of code components as they came in, making it easier to stay on top of changes with lower self-reported cognitive demands. For example, one participant with data science expertise (P3) could recognize the boilerplate data formatting of sample data as it was generated, but found  code for a simulation harder to stay on top of---while another participant with creative coding expertise (P7) found \texttt{p5.js} code more straightforward to retain awareness of.

We observed how participants expecting certain code to come from chat or design panel requests watched as that code streamed in from the {\textsc{ConversationAgent}\xspace} and then engaged in reflection-in-action~\cite{schon_reflective_1984}, expressing surprise (or dismay) when these results deviated from expectations. This behavior echoes the observations by Barke et al.~\cite{barke_grounded_2023} of experienced programmers using GitHub Copilot.

Lastly, some participants (P6, P7, P8) rapidly formed a clear, persistent vision for at least one requested task---and rarely, if ever, found themselves actively seeking design alternatives, feedback, or even an understanding of implicit decisions while in the ``flow state'' of trying to achieve that vision. In the case of P8, the interviewer switched one task's development context from {\textsc{Pail}\xspace} to Anthropic's \textit{Claude} AI system, which could generate and run code that was more aligned with the participant's vision at a more rapid pace than {\textsc{Pail}\xspace}. This approach was an attempt to understand whether P8 would reach a ``saturation point'' where they were satisfied that their vision was achieved, and design support might be welcomed. Though a saturation point was reached, the desired design support was explicitly limited to ``I'm not really interested in what the system might tell me, the only thing I'd want to do is try it with [my child]'' (P8).


\section{Discussion} \label{sec:discussion}

One of our goals in conducting this work is in identifying the next set of challenges the HCI community is likely to face in helping programmers and other technologists design working programs with AI assistance. 
Based on our experience designing {\textsc{Pail}\xspace} and the evaluation results, we identify open challenges for LLM-aided program design around (1) how users will define design goals and problems, \edits{and evaluate progress towards those goals;} and (2) how tools will manage the trade-offs between providing \textit{more complete} information and providing \textit{more relevant} information from the very large set of information that LLMs and other AI systems can inexpensively and rapidly generate. 
These challenges point to a potential shift in the nature of programming, too, with an increased emphasis on \textit{interaction design challenges} like managing user attention and perhaps a decreased emphasis on humans relying on particular \textit{programming language capabilities}---the latter increasingly mediated by the increasing use of higher-level natural language.

\subsection{Defining Goals, Exploring Problem Spaces}

\keyIdea{In order for LLMs to effectively support program design, they must identify steps along the pathway towards a user's ultimate design goal that match a user's mental model of the problem space.} However, as we saw in our evaluation, what steps are meaningful and how important they are depends on where users begin. If users have a point design to begin with, the appropriate next step may be to extrapolate key features before exploring alternatives. In contrast, if users do not know where to begin, a the right next step may be to name a design dimension before exploring concrete instances of it. Indeed, these approaches are characteristic of the double diamond of design as reflected in Figure~\ref{fig:diamonds}.

But though the double diamond implies a certain linearity, we found that {\textsc{Pail}\xspace}'s ability to rapidly generate code along many different directions enabled a very nonlinear approach to design. Participants would rapidly shift between exploring possible solutions and recognizing that their problem formulations may not have been addressable. For LLM-aided tools to be effective in helping users with design, they should facilitate \textit{and recognize} this rapid movement between extrapolation and concretization that is enabled by their code generation capabilities.

In one sense, this reflects a shift from exploratory code as \textit{prototype} to exploratory code as \textit{sketch}. In many contexts, generating running code is considered an engineering project within a ``solution-space'' phase, rather than a ``problem-space'' phase. But, in its role as \textit{sketch}, running code serves a concrete exploratory purpose in evoking what a solution could look like or work like, just as a hand sketch might.

\subsection{Design Practices: Time Compression}

While introducing design-related controls in {\textsc{Pail}\xspace} helped users consider approaches more broadly, {\textsc{Pail}\xspace} sometimes overwhelmed designers with too much \edits{incoming} information with a low signal-to-noise ratio. This risks that users will ignore this information, even when structured to support design. 

Our observations of this overload speak to one possible cause: a design process altered through compression in time by automation. A typical design process in program design relies on individual humans to write code, test applications with users, consider and enumerate alternatives and design rationales---and these activities as a result are paced at a human timescale. In {\textsc{Pail}\xspace}, however, these activities are all accelerated, with agents providing information on multiple facets either concurrently or in quick succession. Given this acceleration, the potential for overload should be clear. One question, then, is how and why designers choose to direct their attention among this accelerated information stream, and what role expertise plays in choosing what to consider carefully, what to skim, and what to ignore.

Such an understanding of designer needs and behaviors would enable future systems to be created with an understanding of what feedback is useful, when, and through what mode of delivery. This is not a new problem: Horvitz identified that a major challenge in automation is the selection of an ``ideal action in light of costs, benefits, and uncertainties''~\cite{horvitz_principles_1999} as early as 1999. 

Ultimately, making sense of the large amount of \textit{potential} information generated by LLM-based agents and other cognitive tools will require a new layer of interaction between human users and the underlying agents producing these insights. The research community is not short on approaches to handling large amounts of data, even when that data changes incrementally over time. But handling large amounts of data that change substantially over time, in contexts where it is hard to assess which of that data is critical to the user, is a challenge that we have only started to consider.

\subsection{Managing ``More Information'' \textit{vs.} ``Better Information''}

It's also not clear who will wield the power \edits{that comes with controlling the information layer}.
Li \textit{et al.}~\cite{li_beyond_2023} have argued that tool designers wield a lot of power to shape thought and practices in the domain of creativity support tools. Vaithilingam  \textit{et al.},~\cite{vaithilingam_imagining_2024}, meanwhile, suggest that LLM-assisted techniques like dynamic grounding can return some of the power to users by allowing tools to \textit{adapt to where humans are}, rather than forcing humans to adapt their ways of thinking and practices to the tools' capabilities as defined by their designers.

Our {\textsc{Pail}\xspace} experiences raise the concern that we are on the brink of handing a lot of power to the IDEs, LLMs, and prompt developers building the next generation of tools, because each new LLM-based design affordance demands attention, and showing them all at once will require a major learning curve. Who directs attention, if not the tool? In a typical creativity support tool, the designers of the tools themselves wield direct control over tool behavior, but in LLM-powered systems, designers often cede varying degrees of control to black-box models they did not even have a hand in training.  We may be moving from a formal system of rules and practices that are at least discoverable and interpretable, to an analog world of prompt-driven components whose very behavior is both inherently unpredictable and also dependent on the model it happens to be executed against. Our experiences with attention overload in other realms (e.g., social media~\cite{orben_windows_2022}) suggest that this future may be less empowering for users and developers, rather than more. 

A key design decision that will shape how systems wield this power goes beyond what information they surface to users and how, and to what information they \textit{generate} and then choose to surface. While a major area of design has been and will continue to be principles for managing information better, our work with {\textsc{Pail}\xspace} also asks if there \textit{better information} to be managing in the first place.


\section{Limitations and Future Work}

\edits{{\textsc{Pail}\xspace} represents one possible point in the design space of AI-supported program design. We did not explicitly compare {\textsc{Pail}\xspace} with ChatGPT or Claude in our evaluation because it was not our goal to show that {\textsc{Pail}\xspace} is more effective than those baselines, but rather to identify the challenges that arise when explicit design support is integrated into an IDE like {\textsc{Pail}\xspace}. We hope our work here opens up a design space with some initial, critical insights about what the community should tackle next.}

\edits{That said, t}here are two key limitations of this work that future work should address: the generalizability of {\textsc{Pail}\xspace} and transitions across levels of abstraction when programming with LLMs. 

First, {\textsc{Pail}\xspace} isn't intended for for every programming task, nor for every programmer. Though {\textsc{Pail}\xspace} helped get participants started regardless of design background, \textit{some} level of design process literacy is required to make effective use of {\textsc{Pail}\xspace}. Design process literacy can be taught, of course---or it can be designed into tools like {\textsc{Pail}\xspace}. For example, recall that participants were most willing to engage with {\textsc{ConversationAgent}\xspace}'s questions about the end-user and their needs when the agent was \textit{not} simultaneously producing changes to the program that a user could be testing---providing code in this context was actively counterproductive. This observation points to a compelling, but also concerning, mechanism by which a tool like {\textsc{Pail}\xspace} could encourage more consideration of various parts of the design process: hold the project hostage by simply refusing to produce any code until the user has considered what the tool wants the user to consider.

Second, regardless of approach, designers need to operate across abstraction levels, knowing when to pop up to a high level from the weeds, and when to deep dive towards a point solution in order to better understand a particular neighborhood within the design space. {\textsc{Pail}\xspace} currently supports \edits{a few levels within its hierarchy of abstraction: code, a layer of established program requirements and decisions embedded in that code, and a set of higher-level design-related concerns.} As recently noted by Vaithilingam \textit{et al.}~\cite{vaithilingam_imagining_2024}, design decisions happen in a fractal pattern, with many decisions across many levels of abstraction. Future LLM-aided programming tools could provide finer-grained control over operation at different levels of abstraction to provide greater control over and understanding of generated code, as well as more targeted legibility for the immediate subtask at hand. We contribute here a deeper understanding that there are substantial challenges designers will face when transitioning across abstraction levels.

Supporting users identifying design problems and solutions, as they handle an information overload and transition across abstraction levels, requires that we tool designers reckon with several major open questions:

How much agency or initiative should automated systems be given to set direction? It's one thing to synthesize some code for a user to evaluate---it's another entirely to control when a user considers their high-level goals for a project instead of staying lost in the weeds, or to scope out a set of potentially-anchoring alternative points in the design space.     
And if programmers spend less time writing code and more time providing higher-level instructions, then managing \textit{attention} in the face of \textit{too much data} and \textit{unknown signal-to-noise ratios} among that data will become critical. How can systems correctly decide what data to show users, and when?

\section{Conclusion}

Through this work, we explored the implications of LLM-aided program design, focused on support for problem formulation and assessing solution suitability. {\textsc{Pail}\xspace}, our design probe, encourages developers to follow a user-centered design process and tracks requirements discovered and decisions made through prototyping. Through our user study, we found some evidence that this kind of assistance can be helpful in broadly considering the program design space, but also uncover a set of challenges around managing attention and maintaining awareness of program updates, pointing to broader questions and trade-offs across generating and sharing information, ensuring information is relevant to users, and balancing agency between users and their new generation of tools.

\section{Disclosure}

The authors used ChatGPT for minor copyediting tasks.

\begin{acks}


The authors would like to thank the many individuals who supported this work, including: Chetan Goenka and Keira Swei, undergraduate researchers supported in their contributions to this work through Berkeley's Engineering Design Scholars summer program; Timothy J. Aveni, Shm Garanganao Almeda, James Smith, and Shreya Shankar, for many fruitful discussions on the concepts underlying this work, as well as feedback on drafts of this paper; our study participants, who graciously gave their time and feedback to this effort; and our anonymous reviewers, whose thoughtful comments shaped many revisions here.

J.D. Zamfirescu-Pereira was partially supported by a Google PhD Fellowship. Qian Yang's effort was supported in part by the AI2050 Early Career Fellowship program at Schmidt Sciences.
This material is also based upon work supported by the National Science Foundation under Grant No. 2313078.
Any opinions, findings, and conclusions or recommendations expressed in this material are those of the author(s) and do not necessarily reflect the views of the National Science Foundation.


\end{acks}

\bibliographystyle{ACM-Reference-Format}
\bibliography{references.bib}

\end{document}